\documentclass{article}

\PassOptionsToPackage{numbers, compress}{natbib}
\usepackage[final]{nips_2016}

\usepackage[utf8]{inputenc} 
\usepackage[T1]{fontenc}    
\usepackage{hyperref}       
\usepackage{url}            
\usepackage{booktabs}       
\usepackage{amsfonts}       
\usepackage{nicefrac}       
\usepackage{microtype}      
\usepackage{graphicx}

\title{Columbia MVSO Image Sentiment Dataset}

\author{
  Vaidehi Dalmia \\
  \texttt{vd2302@columbia.edu}\\
  Columbia University\\
  \And
  Hongyi~Liu\\
  \texttt{hl2906@columbia.edu} \\
  Columbia University\\
  \And
  Shih-Fu Chang \\
  \texttt{sc250@columbia.edu} \\
  Columbia University\\
}

\begin{document}

\maketitle

\begin{abstract}
  The Multilingual Visual Sentiment Ontology (MVSO) consists of 15,600 concepts in 12 different languages that are strongly related to emotions and sentiments expressed in images. These concepts are defined in the form of Adjective-Noun Pair (ANP), which are crawled and discovered from online image forum Flickr. In this work, we used Amazon Mechanical Turk as a crowd-sourcing platform to collect human judgments on sentiments expressed in images that are uniformly sampled over 3,911 English ANPs extracted from a tag-restricted subset of MVSO. Our goal is to use the dataset as a benchmark for the evaluation of systems that automatically predict sentiments in images or ANPs. 
\end{abstract}

\section{Introduction}

The MVSO \cite{jou2015visual} is constructed by collecting images and metadata from Flickr based on emotion keywords from the Plutchik’s wheel of emotions \cite{plutchik1980emotion}. The keywords related to each emotion are used to crawl images and associated metadata, from which frequent, diverse, and sentiment-biased ANPs such as “beautiful flower” or “sad eyes” were discovered. MVSO includes more than 7 million images tagged with these ANPs, with at least 40 images per ANP class.

In this project, we design an AMT interface\footnote{https://www.mturk.com} for annotators to view each image and associate metadata (title and dscription) and assign a sentiment value  between -2 (most negative) to 2 (most positive). Based on our earlier work \cite{borth2013large}, we found sentiment annotations have higher consistency across multiple annotators when both images and textual metadata are shown. We adopt the same strategy in this annotation experiment. In the following, we describe the dataset, the annotation interface, and brief summary of the annotation results. The final sentiment annotation results and the image URL list can be downloaded at the MVSO website\footnote{http://mvso.cs.columbia.edu/download.html}.

\section{Image Dataset} 
We used the MVSO image and metadata dataset for our study \cite{jou2015visual}. The MVSO dataset consists of Flickr images (and metadata such as title and description) that are assigned to different ANPs. Though MVSO consists of ANPs from 12 languages, we only used the English dataset for our image sentiment annotation experiment. To ensure the label quality, in \cite{jou2016going} we also checked whether the ANP name is contained in tags of an image instead of title or description. This filtering process can lead to image-label pairs of higher quality and then more accurate ANP detectors. ANPs with too few tag-restricted images are removed. After filtering, there were 3,911 tag ANPs in English. For each ANP, we randomly sampled 3 images such that the images were uploaded by different users and contained relevant metadata.

\section{Image Sentiment Annotations}
We collected human annotations for our dataset based on sentiment by posting jobs known as Human Intelligent Tasks (HITs) on the Amazon Mechanical Turk crowdsourcing site.

\subsection{Crowdsourcing setup}
\begin{figure}[h]
  \centering
  \includegraphics[width=300pt]{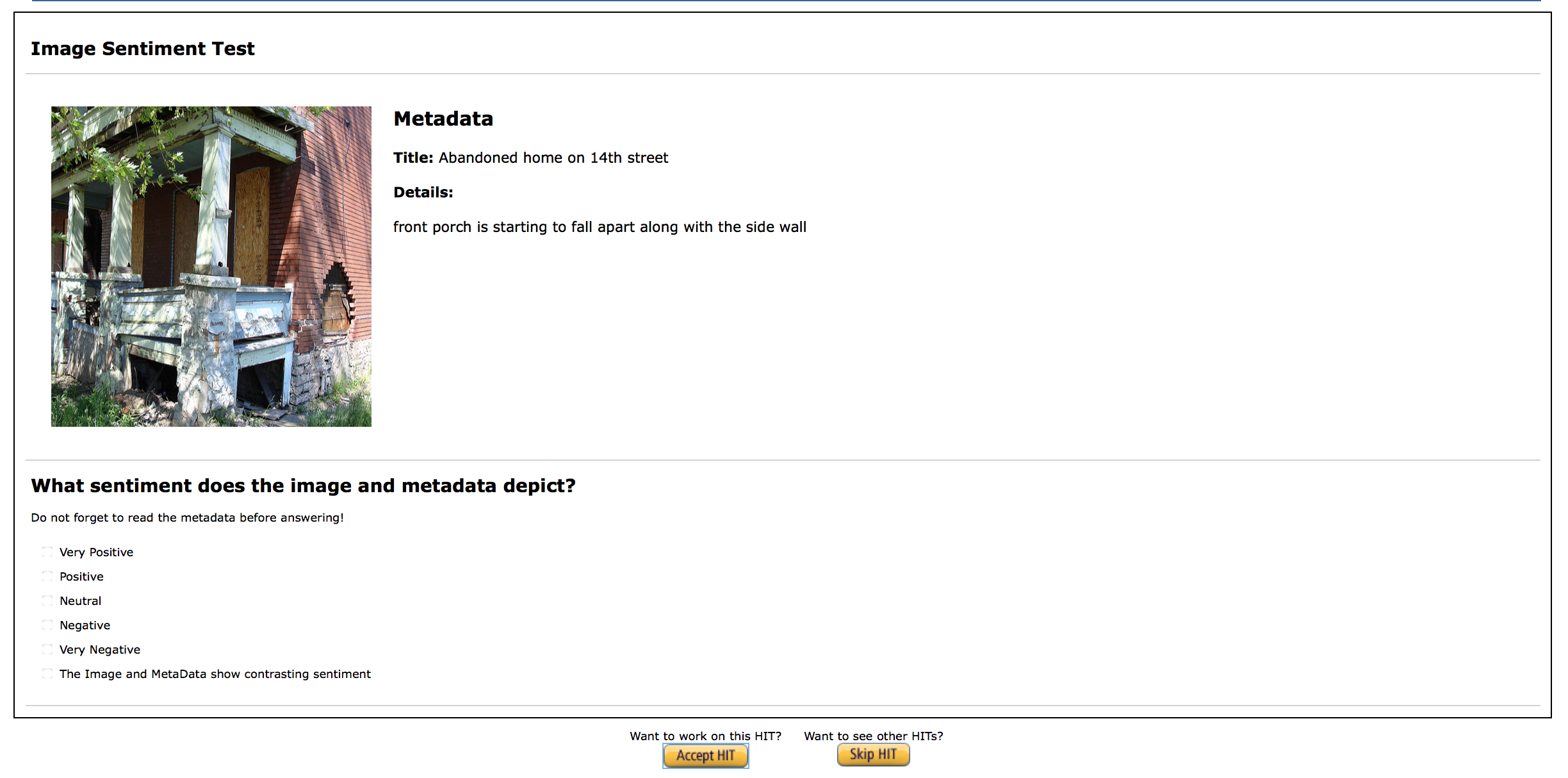}
  \caption{AMT HIT Example. Image from Flickr User: \textit{Jay Carrier}. }
  \label{fig:hit_example}
\end{figure}

We created 3911 $\times$ 3 HITs (3 images per ANP) and had 3 independent workers evaluate each HIT. For the annotation task the workers were asked to look at an image, read the title and description for that particular image, and then based on what sentiment they thought the image and text together conveyed, select one of the following options: very positive, positive, neutral, negative, very negative or contrasting sentiment (between image and text). An example of HIT is shown in Figure. \ref{fig:hit_example}.

\subsection{Quality Control}
\begin{figure}[h]
  \centering
  \includegraphics[width=300pt]{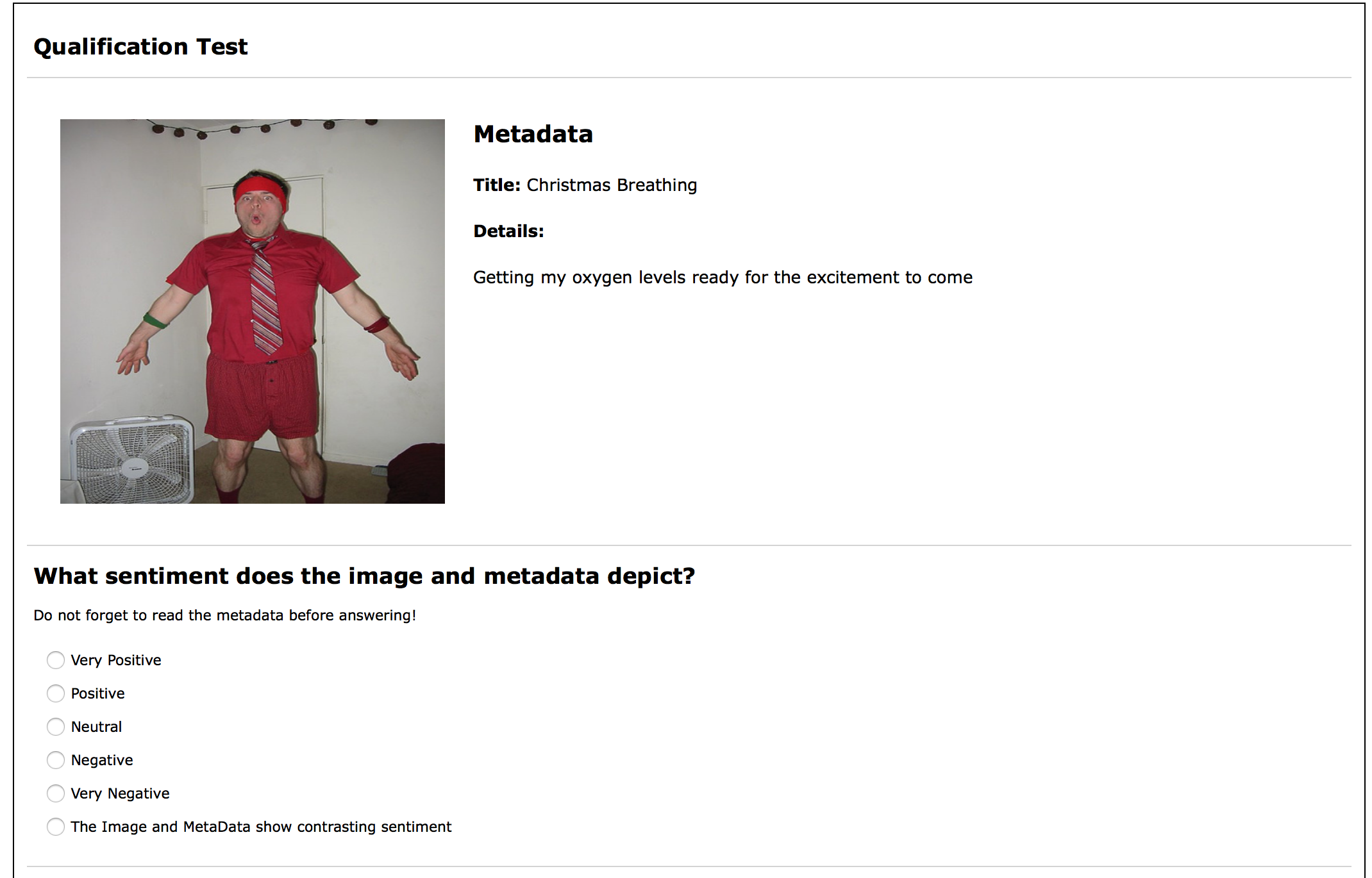}
  \caption{Qualification Test Example. Image from Flickr User: \textit{greggoconnell}. }
  \label{fig:qualification_test}
\end{figure}

Like many other crowdsourcing platforms, Amazon Mechanical Turk provides a quality control mechanism by allowing Requesters to create a Qualification Test. Only workers that have passed the Qualification Test are allowed to access the actual task. For our qualification test, workers were asked 3 questions and had to correctly answer two or more questions right to pass the test. The qualification test questions were carefully picked such that the right answer could only be obtained by both reading the text and looking at the picture. An example of qualication test is shown in Figure. \ref{fig:qualification_test}.

\newpage
\section{Results}
We created three HITs per ANP so the total number of HITs put online was 11,733. Three separate workers or turkers attempted each of these HITs. After all the HITs were evaluated, we collected all the results to look for trends. 
We followed the scoring criterion shown in Table. \ref{tab:scoring_criterion}. Specifically, the sentiment score of a HIT was set as NULL when constrasting sentiment (between image and text) was observed by turkers.
When calculating the standard deviation and sentiment for the images and ANP (described below), the NULL values were ignored. 

\begin{table}[h]
      \caption{Scoring Criterion}
      \small
      \begin{center}
        \begin{tabular}{|c|c|c|c|}
          \hline
          Type & Score & Type & Score\\ \hline
          Very Positive & 2  & Very Negative & -2 \\ \hline
          Positive & 1  &  Negative & -1 \\ \hline
          Neutral & 0  & Contrasting & NULL\\ \hline
        \end{tabular}
        \label{tab:scoring_criterion}\\
      \end{center}
\end{table}

\subsection{Sentiment Analysis per Image}
\begin{figure}[ht]
  \centering
  \includegraphics[width=230pt]{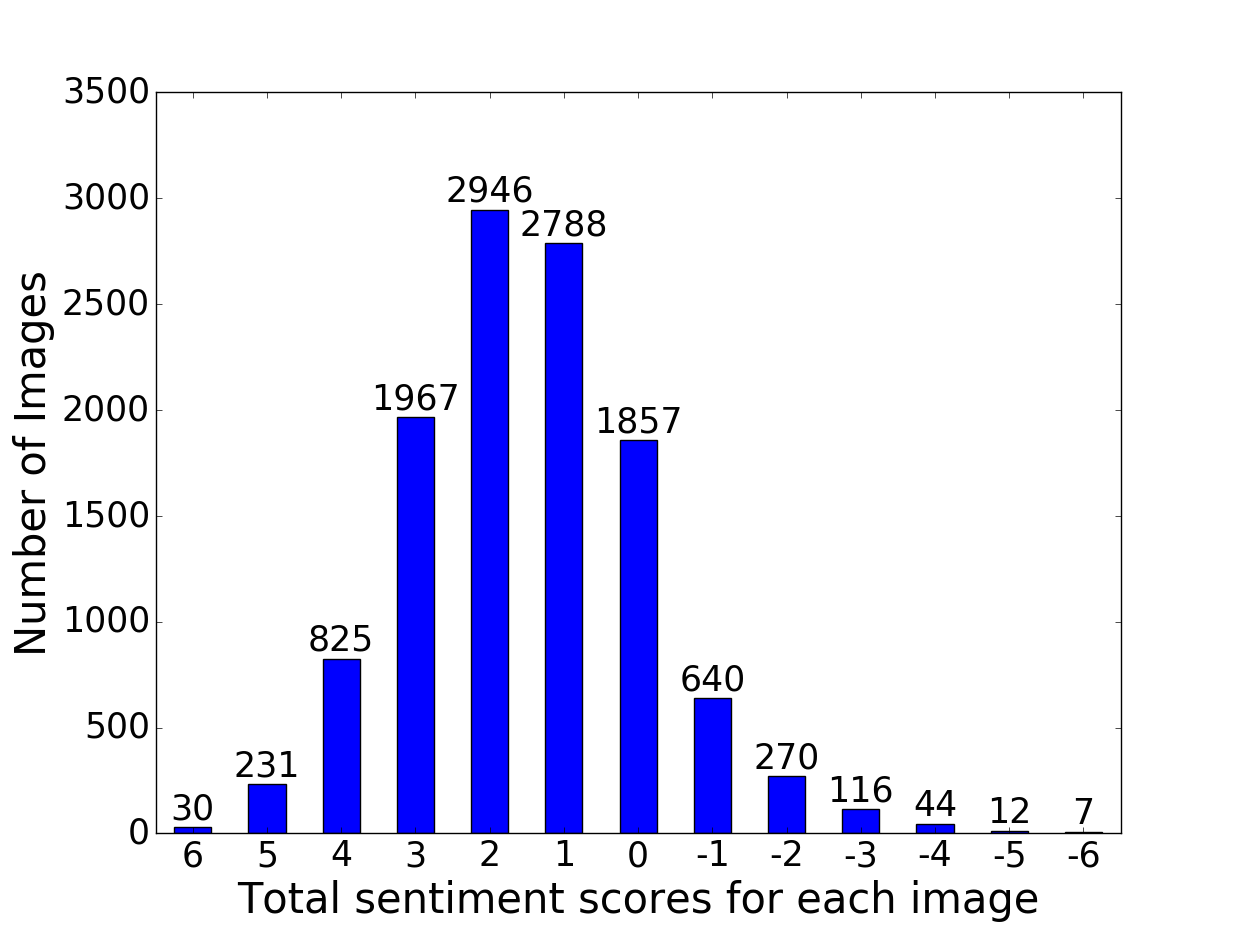}
  \caption{Image count distribution over different sentiment scores}
  \label{fig:image_senti_histo}
\end{figure}

 To see how the sentiment value of our data was distributed across images, we summed the three values we got for each image to get an image sentiment value ranging from -6 to 6. We summed up the number of images with a specific sentiment value and shown the image sentiment distribution in Figure. \ref{fig:image_senti_histo}.

\begin{figure}[h]
  \centering
  \includegraphics[width=150pt]{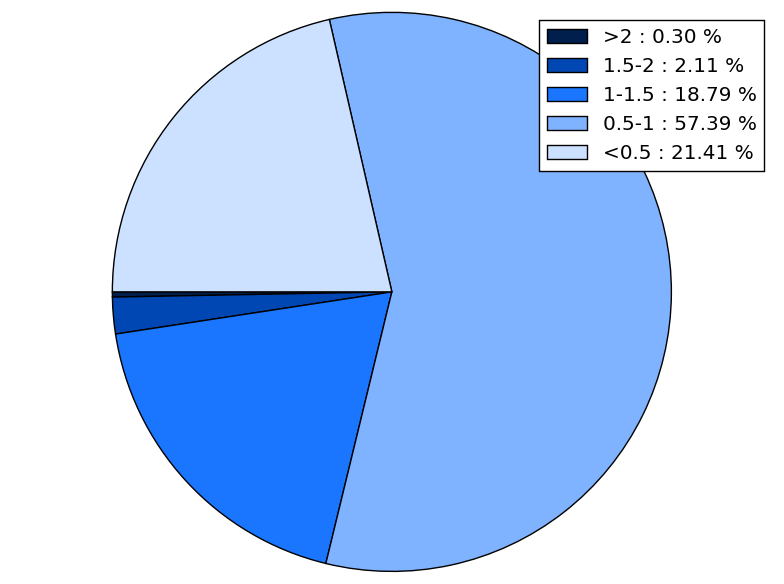}
  \caption{Image Distribution (Standard Deviation)}
  \label{fig:image_senti_std_pie}
\end{figure}

 We also wanted to see how much the values differed among the turkers for each image. Hence, we calculated the standard deviation for each image. Summing up the number of images with standard deviations within certain ranges, we display the standard deviation distribution in Figure. \ref{fig:image_senti_std_pie}.

\subsection{Sentiment Analysis per ANP}
\begin{figure}[h]
  \centering
  \includegraphics[width=300pt]{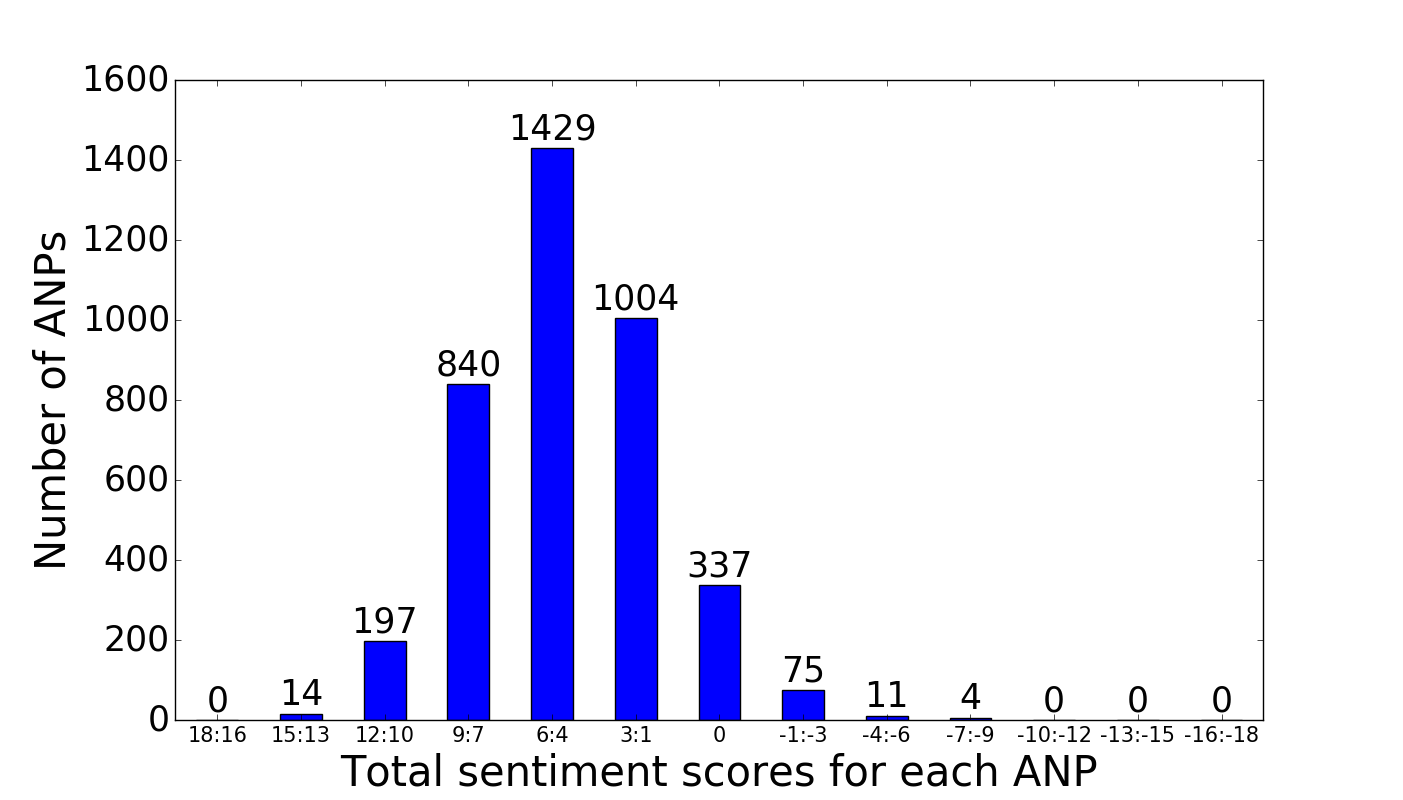}
  \caption{Number of ANPs over sentiment scores}
  \label{fig:anp_senti_histo}
\end{figure}

To see how our sentiment value was distributed across ANPs, we summed the three values we got for each of the three images of an ANP to get an ANP sentiment value ranging from -18 to 18. The distribution of ANPs over summed sentiment values is shown in Figure. \ref{fig:anp_senti_histo}.

\begin{thebibliography}{1}

\bibitem{jou2015visual}
Brendan Jou, Tao Chen, Nikolaos Pappas, Miriam Redi, Mercan Topkara, and
  Shih-Fu Chang.
\newblock Visual affect around the world: A large-scale multilingual visual
  sentiment ontology.
\newblock In {\em Proceedings of the 23rd ACM international conference on
  Multimedia}, pages 159--168. ACM, 2015.

\bibitem{plutchik1980emotion}
Robert Plutchik.
\newblock {\em Emotion: A psychoevolutionary synthesis}.
\newblock Harpercollins College Division, 1980.

\bibitem{borth2013large}
Damian Borth, Rongrong Ji, Tao Chen, Thomas Breuel, and Shih-Fu Chang.
\newblock Large-scale visual sentiment ontology and detectors using adjective
  noun pairs.
\newblock In {\em Proceedings of the 21st ACM international conference on
  Multimedia}, pages 223--232. ACM, 2013.

\bibitem{jou2016going}
Brendan Jou and Shih-Fu Chang.
\newblock Going deeper for multilingual visual sentiment detection.
\newblock {\em arXiv preprint arXiv:1605.09211}, 2016.

\end{thebibliography}

\end{document}